\def\hE{{\displaystyle{\cal E }}}

\def\hJ{{\displaystyle{\cal J }}}

\documentstyle[twocolumn,prl,aps]{revtex}
\begin{document}
\catcode`\@=11
\catcode`\@=12
\twocolumn[\hsize\textwidth\columnwidth\hsize\csname%
@twocolumnfalse\endcsname

\draft
%%%%%%%%%%%%%
\title{Resonant above-threshold ionization 
peaks at quantized intensities}
%%%%%%%%%%%%%
\author{J. Gao$^1$, Dong-Sheng Guo$^2$ 
and Yong-Shi Wu$^{3,\dagger}$}
%%%%%%%%%%%%%
\address{$^1$Department of Electrical Engineering, 
University of Illinois, Urbana, Illinois 61801\\
$^2$Department of Physics, Southern University 
and A\&M College, Baton Rouge, Louisiana 70813 \\
$^3$School of Natural Sciences, Institute for 
Advanced Study, Olden Lane, Princeton, NJ 08540}
%%%%%%%%%%%%%
\date{\today}
%%%%%%%%%%%%%
\maketitle 
\begin{abstract}   
We suggest that electron-laser interactions 
can give rise to resonance phenomena as the 
intensity varies. A new QED perturbation theory is
developed, in which the coupling between an 
electron and the second quantized laser mode 
is treated nonperturbatively. We predict, 
for example, the above-threshold ionization 
rate shows peaks at intensities with integer 
ponderomotive parameter. Such quantum resonance 
effects may be exploited to calibrate laser 
intensities in appropriate range.

\end{abstract}
%%%%%%%%%%%%%
]

\begin{narrowtext}

{\it Introduction} 
\hspace{15pt} For a radiation field, the 
correspondence principle limit requires 
a huge number of photons in unit volume. In 
accordance to this limit, as is widely accepted, 
the light field in a high-intensity laser beam 
can be treated classically or semi-classically. 
Indeed many aspects of the stimulated 
electron-laser interactions have been derived 
in this way \cite{review}. However, according to 
quantum electrodynamics (QED), the fundamental theory 
for electron-photon interactions, even in this 
limit the radiation field still has particle-like 
structure in terms of photons, and stimulated 
emission and absorption give rise to strong
fluctuations in photon number at high light 
intensities. The fluctuations are extremely 
tiny compared to the total photon number, thus 
are ignored in the classical treatments. But 
their {\it absolute magnitude} become appreciable 
at available high intensities and may give rise 
to novel, observable quantum effects.

A physical quantity that can be used to measure 
the strength of stimulated electron-laser-photon 
interactions is the dimensionless {\it 
ponderomotive parameter}:
\begin{equation}
z\equiv U_{\it p}/ \hbar\omega_0=2\pi e^2 
I/m_e \hbar c\omega_0^3, 
 \label{eqno1}
\end{equation}
where $I$ is the intensity, and $\hbar \omega_0$ the 
laser photon energy. It is essentially the 
ponderomotive energy $U_{\it p}$ for an electron in 
the light field in units of the photon energy.
Classically, $U_{\it p}$ is the time-averaged 
kinetic energy of the wiggling motion of the electron 
in response to the radiation field. In quantum 
theory, Eberly first showed\cite{Eber} the 
emergence of the parameter $z$ in stimulated Compton 
scattering. He interpreted it as the number of 
photons in the possibly smallest interaction volume 
for the electron, and argued that when $z$ becomes 
large, many photons at a time are interacting with 
the electron, and stimulated {\it multi-photon 
coherent} emissions and absorptions become frequent 
and significant. We will argue that it is more 
natural to interpret $z$ as the {\it average 
number of laser-mode photons in the photon 
cloud dressing the electron}, whose formation 
is due to stimulated interactions. Thus, when 
the ponderomotive parameter $z$ is close to 
an integer, the dressing photon cloud becomes 
resonant with the laser mode, which has a 
discrete energy spectrum. In view of 
this argument, we suggest that various 
processes in the electron-laser system should 
show {\it resonance-like} structures in their 
{\it intensity} dependence, as the ponderomotive 
parameter $z$ becomes {\it close to an integer}. 

Previously, Guo, \AA berg and Crasemann 
\cite{GAC} have given a nonperturbative 
approach based on almost exact eigenstates 
for the system of an electron coupled to a 
{\it second quantized laser mode}. It was 
then used to derive the above-threshold 
ionization (ATI) rate, leading to nonvanishing 
results only at $z={\rm integer}$. This was a 
sort of precursor of the resonance effects, 
but it could not produce nonzero ATI rate at 
$z\neq {\rm integer}$. To remedy, in 
this paper we develop a new theory for 
QED at high laser photon density, 
by {\it including non-laser radiation 
modes and treat their couplings to the 
electron as perturbation}. In this way we 
are able to demonstrate the resonant ATI peaks 
at $z={\rm integer}$ superposed on a smooth 
backgroud at non-integral $z$. This and 
similar intensity-dependent, resonating 
quantum effects of the light field in other 
multiphoton processes may be experimentally 
exploited to calibrate appropriately 
high laser intensities.

{\it New Perturbation Theory for QED} 
\hspace{12 pt} 
To properly deal with photon number 
fluctuations, we need to {\it quantize 
the electromagnetic field}, but still 
treat the electron quantum mechanically,
ignoring pair production, vacuum 
polarization and other relativistic 
corrections for the electron as well,
if the laser intensity is not too high. 

In the Schr\"{o}dinger picture,
the Hamiltonian of the electron-radiation 
system is 
(with $\hbar=c=1$)
  \begin{equation}
H= {1\over 2m_e}[-i {\bf \bigtriangledown}
- e{\bf A}({\bf r})]^2  
+\sum_{k}\omega_{k} N_{k},
\label{eqno3}
  \end{equation}
with $N_{k}= a^{\dagger}_{k} a_{k} +1/2$.
Here the photon field operator is given 
by the time-independent vector potential
in the radiation gauge 
(${\bf \bigtriangledown} \cdot {\bf A}=0$):
  \begin{equation}
{\bf A}({\bf r}) = \sum_{k} {\bf A}_{k}({\bf r}) 
\equiv \sum_{k} g_{k} ({\bf \epsilon}_{k} 
a_{k} e^{i{\bf k}\cdot{\bf r}}+ {\rm h.c.}),
\label{eqno2}
  \end{equation}
with $k$ labeling the photon modes, including 
the wave vector ${\bf k}$ and transverse 
polarizations described by ${\bf \epsilon}$:
\begin{equation}
{\bf \epsilon}=[{\bf \epsilon}_x 
\cos(\xi/2)+i{\bf \epsilon}_y \sin(\xi/2)]
e^{i\Theta/2}. \\
\end{equation}
Here $g_{k}= (2 \omega_{\bf k} 
V_{\gamma})^{-1/2}$, 
with $\omega_{k}= |{\bf k}|$, and 
$V_{\gamma}$ the normalization 
volume of the radiation field. $a_{k}$ and 
$a^{\dagger}_{k}$ are photon annihilation 
and creation operators.
 
%Because of the significance of multi-photon 
%coherent stimulated emission and absorption 
%in the laser mode, it is necessary to 
Now let us separate the laser modes, say a 
single mode labeled by $k_{0}$, from other 
photon modes: ${\bf A}= {\bf A}_{k_0} + 
{\bf A}'$, and try to first {\it treat the 
electron-laser-mode interactions 
nonperturbatively}, then add the coupling 
of the electron to non-laser modes as 
perturbation. Thus, we are led to split
$H=H_0+V+V'$, with
\begin{eqnarray}
H_0={(-i\bigtriangledown)^2\over 2m_e}
+\omega_0 N_0+\sum_{k'\neq k_0} \omega' N',
\label{parts}\\
V=-{e\over m_e}{\bf A}_{k_0} ({\bf r})
\cdot (-i\bigtriangledown)
+{e^2{\bf A}_{k_0}({\bf r})^2\over 2m_e},
\nonumber\\
V'=-{e\over m_e} {\bf A}'({\bf r})\cdot 
(-i\bigtriangledown) +{e^2{\bf A}_{k_0}
({\bf r})\cdot{\bf A}'({\bf r})\over m_e},
\nonumber
\end{eqnarray}
with the ${\bf A}'^2$ term neglected.

%$H=H_{0}+H'_{0}+H_{1}$ with $H'_{0}= 
%\sum_{k\neq k_{0}}\omega_{\bf k} N_{k}$,
%  \begin{equation}
% H_0= {1\over2m_e}[-i {\bf \bigtriangledown} 
%- e{\bf A}_{k_0}\cdot{\bf r})]^2 
%+\omega_{{\bf k}_0}N_{k_0}
%\label{eqno4}
%  \end{equation}

{}For an electron in the laser field, we 
choose $H_0+V$ as the unperturbed Hamiltonian. 
The eigenstates of $N'$ are simply the Fock 
states for the non-laser mode. For the 
electron-laser-mode subsystem, 
%one may imagine to obtain its eigenstates 
%by starting from the bare state of a free 
%electron with momentum ${\bf p}$ and a Fock 
%state $|n>$ for the laser mode and adiabatically 
%switching on the electron charge $e$ or the 
%laser-mode amplitude $g_{k_0}$. We denote 
%thus obtained eigenstates by 
almost exact eigenstates has been obtained 
before \cite{Guo1}-\cite{Guo2}, which are
labeled by a momentum ${\bf p}$ and an integer
$n$, denoted as $\Psi_{{\bf p}n}^0$. 
They are the nonrelativstic limit of the exact 
solutions\cite{Guo1} to the Dirac equation 
coupled to the quantized laser-mode. They 
form a complete, orthogonal set of states, 
called quantized field Volkov states (QFVS), 
which are the QED analog of the classical 
Volkov states\cite{Volkov}. Their 
non-relativistic limit is verified 
\cite{Guo2} to satisfy the Schr\"{o}dinger 
equation up to errors of the same order 
of magnitude as relativistic corrections. In 
practice, we need only to consider their large 
photon-number limit, $n\rightarrow\infty$, 
$g_{k_{0}}\rightarrow 0$ and 
$\sqrt{n} g_{k_{0}}\rightarrow \Lambda$,
with the QFVS simplified to
  \begin{eqnarray}
\Psi_{{\bf p}n}^0 = V_{e}^{-1/2} 
&& \sum_{j\geq -n} \exp \{i({\bf p}
+(z-j) {\bf k}_{0})\cdot {\bf r}\}\, 
\nonumber\\
&& \quad {\cal J}_{j} (\eta,\zeta_{\bf p},
\phi_{\bf p})^{*}
\exp \{ -ij \phi_{\bf p}\}\,|n+j> .
\label{QFVS}
  \end{eqnarray}
Here $z\equiv e^{2}\Lambda^{2}/m_{e}\omega_{0}$
is the ponderomotive paramater, $\mid n\rangle$ 
a laser-mode Fock state and
  \begin{displaymath}
\eta={1\over 2} z \cos \xi,\;\;
\zeta_{\bf p}={ 2|e|\Lambda\over m_{e}\omega_{0}}
|{\bf p} \cdot {\bf \epsilon}|,
\;\; 
\phi_{\bf p}= \tan^{-1} \bigl(
{p_{y}\over p_{x}} \tan {\xi\over 2}\bigr).
  \end{displaymath}
The ${\cal J}_j$ is compouded from Bessel 
functions $J_m$:
  \begin{equation}
{\cal J}_{j} (\eta,\zeta_{\bf p},\phi_{\bf p})
=\sum_{m=-\infty}^{\infty} J_{m}(\eta)
J_{-j-2m}(\zeta_{\bf p}) e^{2im\phi_{\bf p}},
\label{Bessel}
  \end{equation}
The energy and momentum 
(${\bf P}_{0}=-i{\bf \bigtriangledown}
+N_{k_{0}} {\bf k}_{0}$) eigenvalues 
of the QFVS are given by, respectively,
\begin{eqnarray}
E_{0}({\bf p},n) &=& {\bf p}^{2}/2m_{e} 
+ (n+1/2) \omega_{0} + z \omega_{0}\, ,
\nonumber\\
{\bf P}_{0} ({\bf p},n) &=& {\bf p} 
+ (n+1/2) {\bf k}_{0} + z {\bf k}_{0}\,.
\label{totp}
  \end{eqnarray}

The QFVS is a coherent superposition of
Fock states in the laser mode with 
different photon number; this implies that
the electron in the laser field is dressed 
by a coherent photon cloud which has a 
component in each Fock state with photon 
surplus (or deficit) $j$, generated by
stimulated emission and absorption. We
intepret both the total energy and 
momentum in eq. (\ref{totp}) as consisting 
of contributions from the electron, 
the background photons, and the photon 
cloud, each being on shell. Therefore, it 
is natural to interpret the contributions 
from the photon cloud as the ponderomotive 
energy and momentum \cite{GBG}:
  \begin{equation}
U_{\it p} = z\omega_{0},\,\,\,
{\bf P}_{\it p}=z{\bf k}_{0},
\label{ponder}
  \end{equation}
and {\it identify the average number of 
laser photons in the dressing cloud with 
the ponderomotive parameter} $z$ defined in 
eq. (\ref{eqno1}). We emphasize that 
this interpretation of the ponderomotive
parameter is the distinctive consequence 
of the exact QFVS solutions, not shared 
by any other existing perturbative approaches.

By using the QFVS as unperturbed states, 
we can develop a new perturbation theory
for the electron-radiation system, in 
which the electron-non-laser-mode 
coupling $V'$ is treated as perturbation. 
Then the eigenstate for an electron in
the laser field is the perturbed QFVS, 
$\Psi_{{\bf p}n,n'} = 
\mid {\bf p}n,  n' \rangle
+\mid {\bf p}n, n'\rangle' $, 
with 
\begin{eqnarray}
\mid {\bf p}n, n'\rangle'=
\sum_{{\bf {\tilde p}},{\tilde n},{\tilde n'}}
\mid {\bf {\tilde p}}{\tilde n}, {\tilde n'}
\rangle \, {\langle{\bf {\tilde p}}{\tilde n},
{\tilde n'}\mid V'\mid{\bf p}n,n'\rangle
\over\hE({\bf p}n,n')-
\hE({\bf {\tilde p}} {\tilde n},{\tilde n}')},
\label{prtbst}
\end{eqnarray}
where $\mid {\bf p}n, n'\rangle=
\Psi_{{\bf p}n}^0\mid n' \rangle$, 
with $\mid n' \rangle$ a Fock state 
in a non-laser mode;
$\hE({\bf p}n,n')= E_0 ({\bf p},n)
+(n'+1/2)\omega'$. Note that there is no 
energy shift up to first order. 

%\begin{equation}
%H_I =-{e\over m_e}{\bf A}'({\bf r})
%\cdot (-i\bigtriangledown)
%+{e^2\over m_e} {\bf A_{k_0}}
%({\bf r})\cdot {\bf A}'({\bf r}),
%\end{equation}
%(with the ${\bf A'}^2$ term neglected.

%is transformed into 
%$V_I= {\rm exp}(-iH_0t)H_{I}{\rm exp}(iH_0t)$.
%Then the transition amplitude from 
%$\Phi_{{\bf P}, n}$ to $\Phi_{{\bf P'}, n'}$ 
%up to first order is given by
%  \begin{equation}
%S(({\bf P},n)\rightarrow ({\bf P'},n')) = 
%<\Phi_{{\bf P'},n'}|\Phi_{{\bf P}, n}>\\
%- <\Phi_{{\bf P'},n'}|V_I|\Phi_{{\bf P}, n}> . 
%\label{eqno5}
%  \end{equation}

%We start with a proper description of an electron,
%in a single-mode laser field, dressed by a photon 
%cloud due to stimulated emission and absorption.
%Such a state is described by the so-called
%QED Volkov state, first introduced by one of us and
%{\AA}berg\cite{Guo1}, which is the eigenstate of the
%electron-laser-mode system, with the (nonrelativistic)
%electron first quantized and the laser mode second 
%quantized.

%\vspace{4pt} 
{\it Calculation of the ATI Rate} 
\hspace{30pt} 
{}For definiteness, let us consider the 
above-threshold ionization (ATI): A beam of 
neutral atoms with tightly bound electrons is 
injected into a monochromatic, elliptically 
polarized single-mode laser beam. Even if
the photon energy is much less than the 
ionization energy, a bound electron can
absorb simultaneously quite a number of, 
say ten to twenty, photons to become 
ionized with appreciable kinetic energy.
Of course, before the ionized electron
gets out of the laser beam, it has 
very strong stimulated interactions
with the laser mode. During the ATI process, 
to balance the total energy and momentum, 
the electron may emit a photon 
not in the laser mode (spontaneous emission).
We want to calculate the rate and angular 
distributions etc., and study their 
intensity dependence. 

Let us start with the following initial 
state for the elec-\\tron-radiation system: 
the electron in a bound state 
$\Phi_i$, the laser mode in the Fock 
state $\mid n_i\rangle$, 
and the non-laser modes in the vacuum 
state (with $n_i'=0$ photons), denoted
by $\mid \Phi_i, n_i, 0\rangle$. In 
the final state of the ATI, denoted
as $\mid {\bf P}_f, n_f, n_f'\rangle$,
the electron is in a free state with 
momentum ${\bf P}_f$ outside the laser 
beam, the laser mode in the state 
$\mid n_f\rangle$ and at most one, 
say $k'$, of the non-laser modes in 
$\mid n_f'=1\rangle$ (to first order). 
Physically the ATI can be viewed as 
a two-step process \cite{GAC}:
1) The electron is first ionized into 
the laser field, so the intermediate 
state of the system is described by the 
perturbed QFVS given by eqs. (\ref{QFVS})
and (\ref{prtbst});
2) The electron exits out of the laser
beam and becomes a free electron. 
%To simplify, we will assume that the 
%laser beam has a sharp edge, so that 
%the exiting happens suddenly.

With this physical picture in mind, we
apply the standard formal theory for
scattering \cite{GL} and its adaption
to the present situation \cite{GAC},
up to first order in perturbation 
theory, resulting in the transition 
amplitude
\begin{eqnarray}
T_{fi}=\sum_{{\bf p},n, n'}
&&\langle{\bf P}_f,n_f,n_f'
\mid \Psi_{{\bf p}n,n'}\rangle
\nonumber\\
&&\qquad\langle\Psi_{{\bf p}n,n'}
\mid V + V'\mid\Phi_i,n_i,0\rangle,
\label{trans}
\end{eqnarray}
where the summation of intermediate 
states is subject to 
%constrained by the condition that
\begin{eqnarray}
\hE ({\bf p}n,n')
=\hE_i\equiv -E_b+(n_i+{1\over2})
\omega_0+{1\over2}\omega',
\nonumber\\
=\hE_f\equiv{\bf P}_f^2/2m_e +(n_f+1/2)
\omega_0+(n'_f+1/2)\omega'\, ,
\label{cons}
\end{eqnarray}
with $E_b$ the binding energy in the 
initial state $\Phi_i$, while both 
$n'$ and $n_f'$ are either 0 or 1, up 
to first order.
%spontaneously emitted mode, its free photon 
%number $n_f'$ can only be 1 or 0. 

%$\hE_\mu$ stands for the energy of 
%either perturbed or unperturbed QFVS, 
%while $\hE_f$ is actually the energy 
%level of the ATI final state. 
%$\mid{\bf P}_f,n_f,n_f'\rangle$. 

%\begin{eqnarray}
%\Psi_f=\sum_{\mu}\mid \Psi_\mu^0
%+ \Psi_\mu'\rangle\langle\Psi_\mu^0+ 
%\Psi_\mu'\mid{\bf P}_f,n_f,n_f'\rangle,
%\label{final}
%\end{eqnarray}
%where $\mu=({\bf P},n)$, and $\hE_\mu
%=\hE_f\equiv{\bf P}_f^2/2m_e
%+(n_f+1/2)\omega+(n'_f+1/2)\omega'$.

Inspection shows only the following 
terms are nonzero:
\begin{eqnarray}
&&T_0=\sum_{{\bf p},n}
\langle{\bf P}_f,n_f,0\mid {\bf p}n,0\rangle
\langle{\bf p}n,0 \mid V \mid\Phi_i,n_i,0\rangle,
\label{T0}\\
%=T^{\rm GAC},\\
&&T_1=\sum_{{\bf p},n}\langle{\bf P}_f,n_f,1
\mid {\bf p}n,1\rangle \langle {\bf p}n,1\mid 
V '\mid\Phi_i,n_i,0\rangle,\nonumber\\
&&T_2=\sum_{{\bf p},n}\langle{\bf P}_f,n_f,1
\mid {\bf p}n,1 \rangle\langle {\bf p}n,1\mid' 
V\mid\Phi_i,n_i,0\rangle,\nonumber\\
&&T_3=\sum_{{\bf p},n}\langle{\bf P}_f,n_f,1
\mid {\bf p}n,0 \rangle' \langle {\bf p}n,0 
\mid V\mid\Phi_i,n_i,0\rangle.
\label{Ts}
\end{eqnarray}
The zeroth order term $T_0$ has been calculated 
before\cite{GAC}. $T_1$ and $T_2$, as well as
$T_0$, contribute only at $z= {\rm integer}$, 
while $T_3$ contributes both at $z= {\rm integer}$ 
and $z\ne {\rm integer}$. We are interested in 
$z\ne {\rm integer}$, so we focus on $T_3$. 

After a lengthy calculation, introducing 
$j=n_i-n$, $j'=n_f-{\tilde n}$ and $q=n_i-n_f$,
we finally obtain 
\begin{eqnarray}
T_3&&={eg'\omega_0\over m_e}V_e^{-1/2}\Phi_i({\bf P}_f
-q{\bf k}+{\bf k}')e^{iq\Theta/2}{1\over\omega_0} 
\sum_{j,j'}{j-z\over z-j'}\nonumber\\
&&\hJ_{j'}^*(\zeta_{{\bf P}_f}},\eta, \phi_{{\bf P}_f})
e^{-ij'\phi_{{\bf P}_ f}}
\hJ_j(\zeta_{{\bf P}_f+{\bf k}'},\eta, 
\phi_{{\bf P}_f+{\bf k}'})
e^{ij\phi_{{{\bf P}_f+{\bf k}'}}\nonumber\\
&&[-({\bf P}_f+(j-q-z){\bf k})\cdot{{\bf \epsilon}}'^* 
J_{q-j+j'}(\zeta_{{\bf k}'})e^{i(q-j+j')\phi_{{\bf k}'}}
\nonumber\\
&&+e\Lambda{{\bf \epsilon}}^*\cdot{{\bf \epsilon}}'^*
J_{q-j+j'+1}(\zeta_{{\bf k}'})
e^{i(q-j+j'+1)\phi_{{\bf k}'}+i\Theta/2}\nonumber\\
&&+e\Lambda{{\bf \epsilon}}\cdot{{\bf \epsilon}}'^*
J_{q-j+j'-1}(\zeta_{{\bf k}'})
e^{i(q-j+j'-1)\phi_{{\bf k}'}-i\Theta/2}].
\label{T3}
\end{eqnarray}

A careful study shows that the 
kinetic energy difference for the 
photoelectron before and after exiting 
out of the light field is of the order 
of relativistic corrections.
Thus, energy conservation implies a 
discrete spectrum for both
the photoelectron and the non-laser 
photon: 
\begin{eqnarray}
\omega'&&\approx [z-(j-q)]\omega_0,
\nonumber\\
{\bf P}_f^2/ 2m_e &&\approx
j\omega_0-E_b-z\omega_0\ge 0.
\label{26}
\end{eqnarray}
The physical interpretation is clear: the 
electron is ionized by absorbing $j$ photons 
simultaneously and, upon exiting out of the 
laser field, completely shakes off its 
ponderomotive energy (or the dressing photon 
cloud), by emitting $j-q$ laser photons and 
a non-laser photon with the remaining 
ponderomotive energy.
 
We express the energy delta function 
$\delta(\hE_i-\hE_f)$ as 
\begin{eqnarray}
%\delta(\hE_i-\hE_f)=
({m_e\over2\omega_0})^{1/2}
{\delta[P_f-(2m_e\omega_0)^{1/2}
(q-\epsilon_b-\nu)^{1/2}]\over
(q-\epsilon_b-\nu)^{1/2}},
\label{27}
\end{eqnarray}
where $\epsilon_b\equiv E_B/\omega_0$ 
and $\nu\equiv \omega'/\omega_0$ and 
$P_f=\mid{\bf P}_f\mid$.
Then the total ATI rate is given by 
\begin{eqnarray}
W=\int_6 {V_eV_{\gamma}\over(2\pi)^6}\mid 
T_3 \mid^2 2\pi \delta(\hE_i-\hE_f)
d^3{\bf P}_fd^3{\bf k}_f',
\end{eqnarray}
while angular distribution is given by
\begin{eqnarray}
&&{d^5W\over d\Omega_{{\bf P}_f} 
d\Omega_{{\bf k}'} d\omega'}
={e^2\omega'\over 2m_e^2 (2\pi)^3}
{(2m_e^3\omega_0^5)^{1/2}\over(2\pi)^2}
(q-\epsilon_b-\nu)^{1/2}\nonumber\\
&&\qquad \qquad \mid\Phi_i({\bf P}_f-q{\bf k}
+{\bf k}')\mid^2
\mid{\hJ}_{q}^*({\bf P}_f,{\bf k}')\mid^2\, .
\end{eqnarray}
where ${\hJ}^*_q({\bf P}_f,{\bf k}')$ is
the sum $(1/\omega_0)\sum_{j,j'}$ in eq. 
(\ref{T3}).

\vspace{3pt}
{\it Resonant ATI Peaks}\hspace{30pt}
{}From eq. (\ref{T3}) one can infer 
that the amplitude $T_3$ becomes very
large, if $z$ is sufficiently close 
to an integer, because then one of the 
terms in the sum can have a very small 
denominator $j'-z$. Thus, we predict 
that there are resonant ATI peaks at 
quantized intensities with $z=N$ or
\begin{equation}
I= N I_0 \equiv N {\hbar m_e c 
\omega_0^3 \over 2\pi e^2},
\label{peaks}
\end{equation}
with $N\geq 1$ an integer. Note that 
$I_0$ is proportional to the cube of 
the laser frequency $\omega_0$.

As example, in Fig. 1 we present the 
numerically calculated total 
photoelectron counts, collected in the 
direction of polarization, for ATI of 
xenon in a single-mode linearly polarized 
laser beam with wavelength 1064 nm. It 
indeed confirms the emergence of resonance 
peaks at quantized intensities, with 
$I_0=1.10\times10^{13}$ W-${\rm cm}^{-2}$.
%as predicted.
Note the smooth background away from the 
resonant peaks, grossly dictated by 
the classical description of the laser 
field. Becasue the widths of the 
resonance peaks in Fig. 1 are rather 
narrow, for the ATI measurements to 
test our theory the laser intensity 
should be very stable and adjustable 
almost continuously.

To compare, we recall that the 
Keldysh-Faisal-Reiss theory \cite{KFR} 
gives the ATI amplitude
\begin{eqnarray}
T_{fi}^{KFR}=
\langle {\bf P}_f\mid V\mid \Phi_i\rangle,
\end{eqnarray}
where the final state is the Volkov state
in a classical light field. Unlike our $T_3$,
it neither predicts the resonance peaks nor
emission of a non-laser photon $\omega'$.

In the above, as in the usual treatments, 
we have ignored the effects of the 
ionic Coulomb interactions, since 
the Coulomb matrix elements 
are of the order $e^2$ while those of 
$V'$ are of the order $e$. We have also 
used the Fock states as the basis for 
the laser mode. If one uses Glauber
coherent states to describe the initial 
and final states of the laser field, the 
ATI amplitude can be easily derived by
superposing our amplitudes. This gives 
rise to a spread in photon number $n_i$ 
and $n_j$. But the corresponding spread 
in $z$ is expected to be very small. So 
our prediction of the resonant ATI 
peaks is unaffected. 

Our ATI rate diverges at exactly integral
$z$. This problem is easy to remedy, by 
including in eq.(\ref{prtbst}) an imaginary 
part (a finite width) for the QFVS energy 
$\hE({\bf {\tilde p}} {\tilde n},0)$, 
due to its ability to decay through 
spontaneous emission via the coupling $V'$ 
to non-laser modes.

{\it Other Intensity-Dependent Quantum 
Effects} \hspace{15pt}
Our argument for the resonance effects 
in the electron-laser system is 
very general, based only on the intensity 
dependent stimulated interactions and 
the discrete photon structure of the 
laser mode. So we expect to see them
in other multiphoton processes, and our 
approach to QED at high laser photon 
density is applicable as well.

One example is a slow electron transversing 
a single-mode laser beam. Classically, the 
ponderomotive energy acts like an effective 
repulsive potential, so at high intensities
the electron can hardly get into the laser beam. 
But according to our argument, the stimulated 
electron-laser interactions will give rise to 
a photon cloud dressing the electron, which 
can be resonant with the laser mode. So we 
predict that when the laser intensity is 
close to the quantized values $N I_0$, there 
will be resonance peaks for the penetration 
probability for slow electrons transversing 
the laser beam. Our new perturbation theory 
is applicable to make quantitative predictions.

It is easy to generalize our approach to 
more than one laser modes, since the 
corresponding QFVS have been obtained 
before \cite{Guo4}. For example, one may 
consider electrons scattered by a standing 
wave formed by two laser modes. Previously, 
Bucksbaum {\it et al.} \cite{Buck} has 
experimentally discovered a dramatic peak 
splitting in the angular distribution of the 
scattered electron. This has been theoretically 
explained in ref. \cite{GD} using the QFVS 
states, which could not deal with the angular 
region inside the splitting angle. Our new 
perturbation theory can be employed to deal 
with the angular region in between the peaks, 
and is expected to reveal a characteristic 
variation in the peak separation as $I/I_0$ 
changes continuously near an integer.

These and similar intensity-dependent 
quantum effects of the light field, if 
verified, would provide means for 
calibrating the intensity of laser beams 
in the range with the order of $I/I_0$ 
from unity to at least few tens. 

{\it Acknowledgements} \hspace{15pt}
J.G. thanks Prof. J.G. Eden for support and
discussions. D.S.G. was supported in part
by NSF grant No. PHY-9603083, Y.S.W. 
by NSF grant No. PHY-9601277 and a grant 
from Monell Foundation.

\begin{figure}
\caption{Total ATI photoelectron count, collected 
in the laser polarization direction, vs. laser 
intensity for xenon in a single-mode, linearly 
polarized laser at 1064 {\it nm}.} 
\label{fig1}
\end{figure}

\end{narrowtext} 
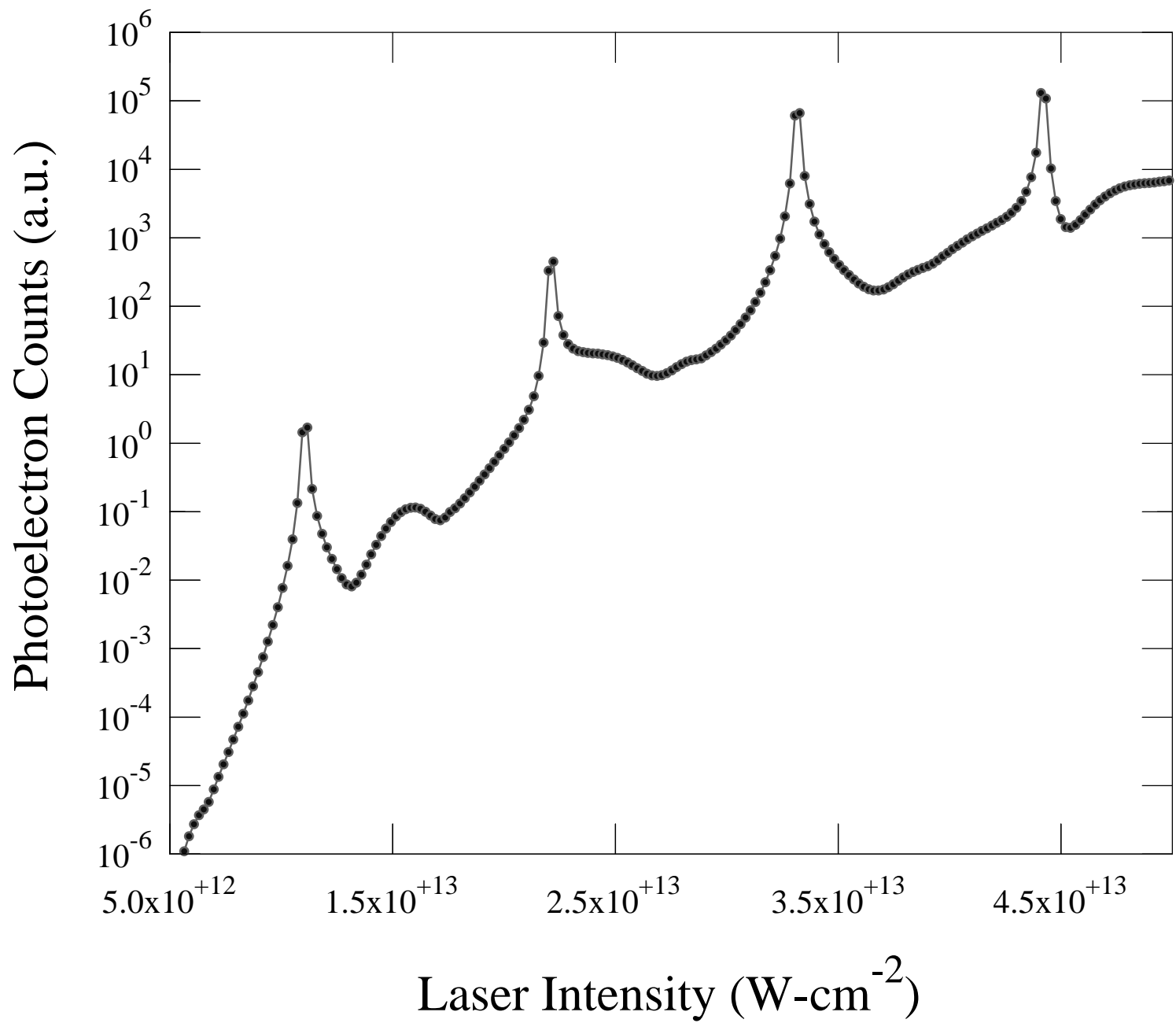
\end{document}